\documentclass[journal]{IEEEtran}

\usepackage{graphicx}             
\usepackage{amsmath}              
\usepackage{amssymb}              
\usepackage[caption=false, font=footnotesize]{subfig}
\usepackage{cite}
\usepackage{siunitx}
\usepackage{bm}

\usepackage[nomargin,inline,draft,mode=multiuser,author=]{fixme}
\fxusetheme{colorsig}
\definecolor{fxwarning}{rgb}{0.8,0.0000,0.0000}
\FXRegisterAuthor{tfl}{atfl}{TFL}
\FXRegisterAuthor{wz}{awzl}{WZ}

\usepackage[capitalize]{cleveref} 

\renewcommand\vec{\bm}
\newcommand{\m}[1]{\left(\begin{matrix}#1\end{matrix}\right)}
\newcommand{\s}[1]{\ \si{#1}}

\title{Real-Time Blind Photonic Interference Cancellation for mmWave MIMO}
\author{
	\IEEEauthorblockN{
		Joshua~C.~Lederman\IEEEauthorrefmark{1},
		Weipeng~Zhang\IEEEauthorrefmark{1},
		Thomas~Ferreira~de~Lima\IEEEauthorrefmark{2}\IEEEauthorrefmark{1},
		Eric~C.~Blow\IEEEauthorrefmark{1},
		Simon~Bilodeau\IEEEauthorrefmark{1},
		Bhavin~J.~Shastri\IEEEauthorrefmark{3},
		Paul~R.~Prucnal\IEEEauthorrefmark{1}}\\
	\IEEEauthorblockA{\IEEEauthorrefmark{1}Department of Electrical and Computer Engineering,
	Princeton University, Princeton, NJ 08544, USA\\}
	\IEEEauthorblockA{\IEEEauthorrefmark{2}NEC Laboratories America, Princeton, NJ 08540, USA\\}
	\IEEEauthorblockA{\IEEEauthorrefmark{3}Department of Physics, Engineering Physics \& Astronomy, Queen’s University, Kingston, ON K7L 3N6, Canada.\\}

}

\date{\today}

\begin{document}
\maketitle
\begin{abstract}

	Multiple-input multiple-output (MIMO) mmWave devices broadcast multiple spatially-separated data streams simultaneously in order to increase data transfer rates. Data transfer can, however, be compromised by interference. Conventional techniques for mitigating interference require additional space and power not generally available in handheld mobile devices. Here, we propose a photonic mmWave MIMO receiver architecture capable of interference cancellation with greatly reduced space and power needs. We demonstrate real-time photonic interference cancellation with an integrated FPGA-photonic system that executes a novel zero-calibration micro-ring resonator control algorithm. The system achieves sub-second cancellation weight determination latency with sub-Nyquist sampling. We evaluate the impact of canceller design parameters on performance, establishing that effective photonic cancellation is possible in handheld devices with less than 30 ms weight determination latency. 

\end{abstract}

\section{Introduction}

\begin{figure}
	\centering
	\includegraphics[width=\linewidth]{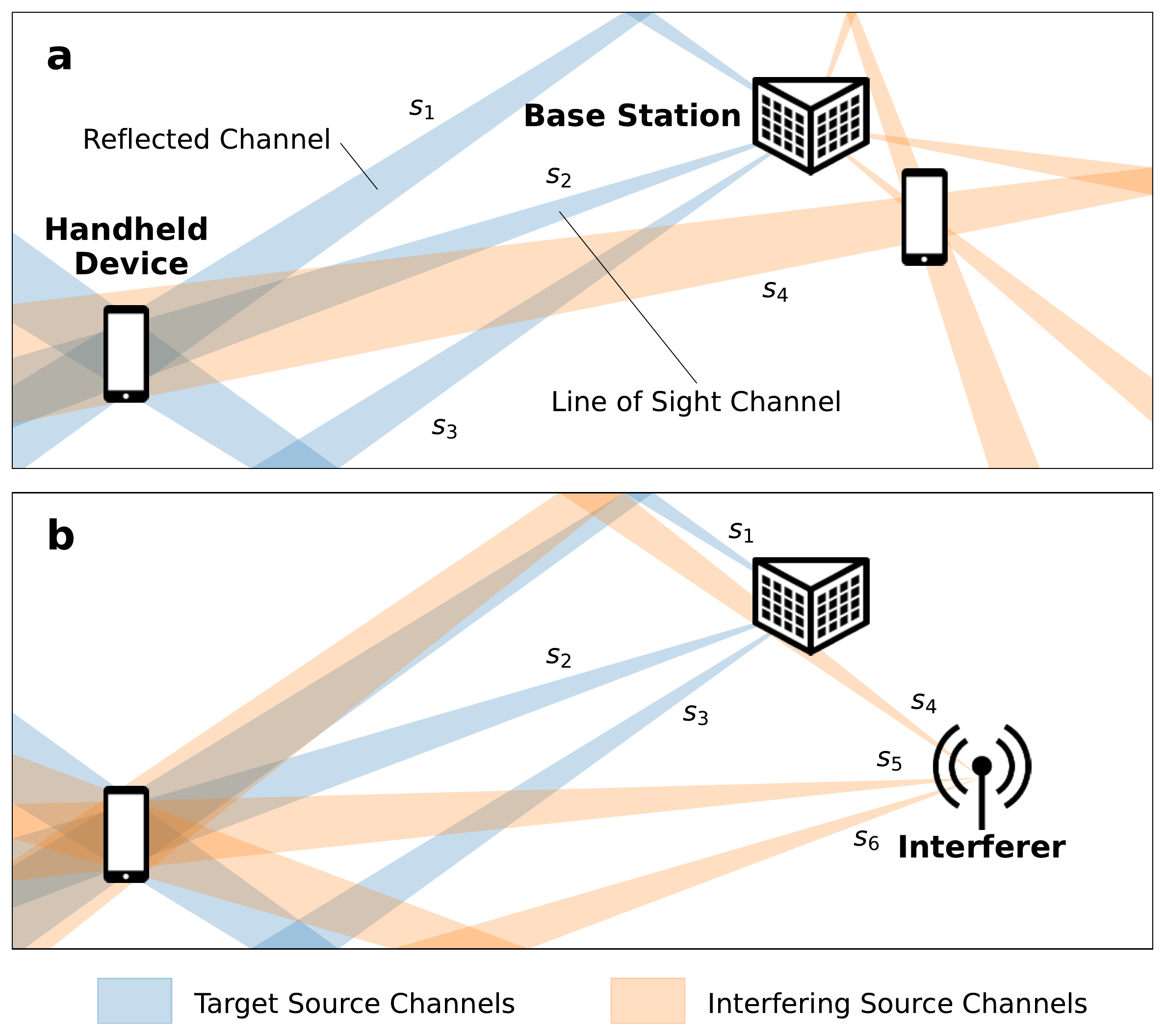} 
	\caption{mmWave interference scenarios. (a) A spatial channel associated with one device interferes with another device's spatial channels.
	(b) A malicious source directs interference at a device.}
	\label{fig:task}
\end{figure}

\noindent Technological development has driven an ever-increasing demand for wireless communication bandwidth\cite{nauman_multimedia_2020, shafique_internet_2020, watanabe_2021}. With sub-7 GHz bands heavily utilized, the industry has turned to 24-53 GHz (mmWave) bands to meet these growing needs \cite{alkhateeb_2014, kutty_beamforming_2016}. At these frequencies, further capacity gains can be realized using beamforming---the angular steering and filtering of radio-frequency (RF) signals to optimize wireless communication links. mmWave beamforming may be implemented using an array of mm-scale antennas in one package suitable for a handheld device \cite{watanabe_2021, alkhateeb_2014, sohrabi_2016, kutty_beamforming_2016}. Hybrid analog-digital beamformers are capable of transmitting or receiving signals at multiple angles simultaneously, allowing independent data streams to be transmitted along different spatial paths at the same time (\cref{fig:task}) \cite{watanabe_2021, abbas_millimeter_2017, alkhateeb_2014, alkhateeb_2015, kutty_beamforming_2016}. This multiple-input multiple-output (MIMO) approach multiplies the capacity of an RF link \cite{watanabe_2021, alkhateeb_2014, sohrabi_2016, kutty_beamforming_2016}. 

\begin{figure*}
	\centering
	\includegraphics[width=\linewidth]{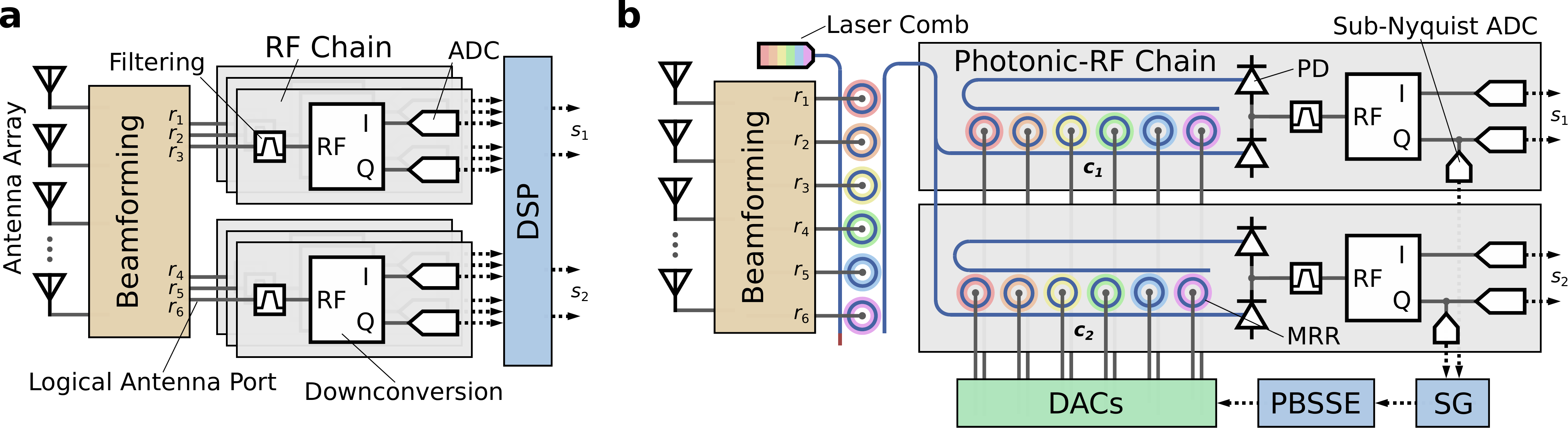}
	\caption{mmWave beamforming receivers. (a) A conventional hybrid beamforming receiver. (b) A photonic-electronic hybrid beamforming receiver. Each receiver includes a beamforming phase-shifter array linking the antenna array to six logical antenna ports, which provide signals ($r_1, r_2, \ldots r_6$). The conventional receiver requires six RF chains to digitize the in-phase (I) and quadrature (Q) components of all sources, with digital interference cancellation allowing the extraction of target source signals $s_1$ and $s_2$. The photonic-electronic receiver performs interference cancellation using RF photonics prior to downconversion, and only two photonic-RF chains are required for the two target signals. Sub-Nyquist analog-to-digital converters (ADCs) provide samples from the target Q signals to drive the PBSS algorithm. DSP: Digital Signal Processing, PD: Photo-Detector, PBSSE: PBSS Engine, SG: Statistic Generator.}
	\label{fig:diagram}
\end{figure*}

Beamforming receivers are imperfect, receiving signals from some off-target angles, particularly those close to the angles of target signals. Interfering signals arriving at such angles introduce noise that degrades network capacity \cite{kutty_beamforming_2016, dai_2015, heath_2016, jensen_review_2004, krishnaswamy_2016}. This interference may stem from the other spatial channels, including those of other devices on the network using the same spectral resources as in a multi-user MIMO system (\cref{fig:task}a) \cite{alkhateeb_2014, kutty_beamforming_2016}, or from a malicious actor (\cref{fig:task}b).

Interference may be mitigated by isolating each incoming signal according to its angle of incidence and subtracting it from the target signals, improving their signal-to-noise ratios (SNRs) \cite{kutty_beamforming_2016, dai_2015, albreem_2019, ye_1998, zhan_2021}. When implemented digitally, this technique requires an independent RF chain for every interference source in order to generate the associated digital reference signal. mmWave RF chains are costly in space and power, limiting their quantity in handheld devices, where these resources are constrained \cite{abbas_millimeter_2017, alkhateeb_2014, krishnaswamy_2016, skrimponis_2020}. Analog interference cancellation, where the interference is subtracted prior to signal digitization, addresses this limitation. However, it requires precise amplitude control, and the variable-gain amplifiers used for analog RF amplitude control face significant limitations related to circuit size and complexity, gain resolution, and consistent performance across frequencies and gain levels \cite{zhao_2019, zhang_ka_2020}. Instead, we propose implementing mmWave MIMO interference cancellation using RF photonics.

CMOS-compatible integrated silicon photonic devices have proven effective at broadband, tunable, high-precision scaling of RF signals modulated onto optical carriers \cite{weigel_bonded_2018, salamin_plasmonic_2018}. A single micro-ring resonator (MRR) has achieved 9 bits of amplitude weight precision and effective control of signals from DC to 19.2 GHz using only 1500 \si{\micro\meter\squared} of chip area \cite{zhang_2023}. Using the broadcast-and-weight architecture \cite{tait_broadcast_2014}, a compact photonic system can perform broadband weighting and summing of RF signals, enabling photonic self-interference cancellation \cite{chang_2017, blow_2018}, and photonic blind source separation (PBSS) \cite{tait_2018, ma_photonic_2019, ma_2020, zhang_2021, huang_2022, zhang_2023}.

In this work, we present a novel photonic mmWave beamforming receiver architecture capable of broadband interference cancellation with simpler hardware and lower power requirements than conventional alternatives. Our proposed approach relies on \emph{real-time} photonic weight determination---weight determination with a latency small compared to the timescale of environmental change---which has not previously been reported. We demonstrate real-time photonic interference cancellation, made possible by two innovations: integration of an MRR weight bank with a field-programmable gate array (FPGA) and the development of a zero-calibration MRR control scheme. We achieve reliable sub-second cancellation weight determination latency with sub-Nyquist sampling, features critical to real-world operation of the proposed system in a handheld mobile device. We further establish that despite tradeoffs between hardware complexity, latency, and accuracy, effective photonic interference cancellation is possible with less than 30 ms weight determination latency.

\begin{figure*}
	\centering
	\includegraphics[width=\linewidth]{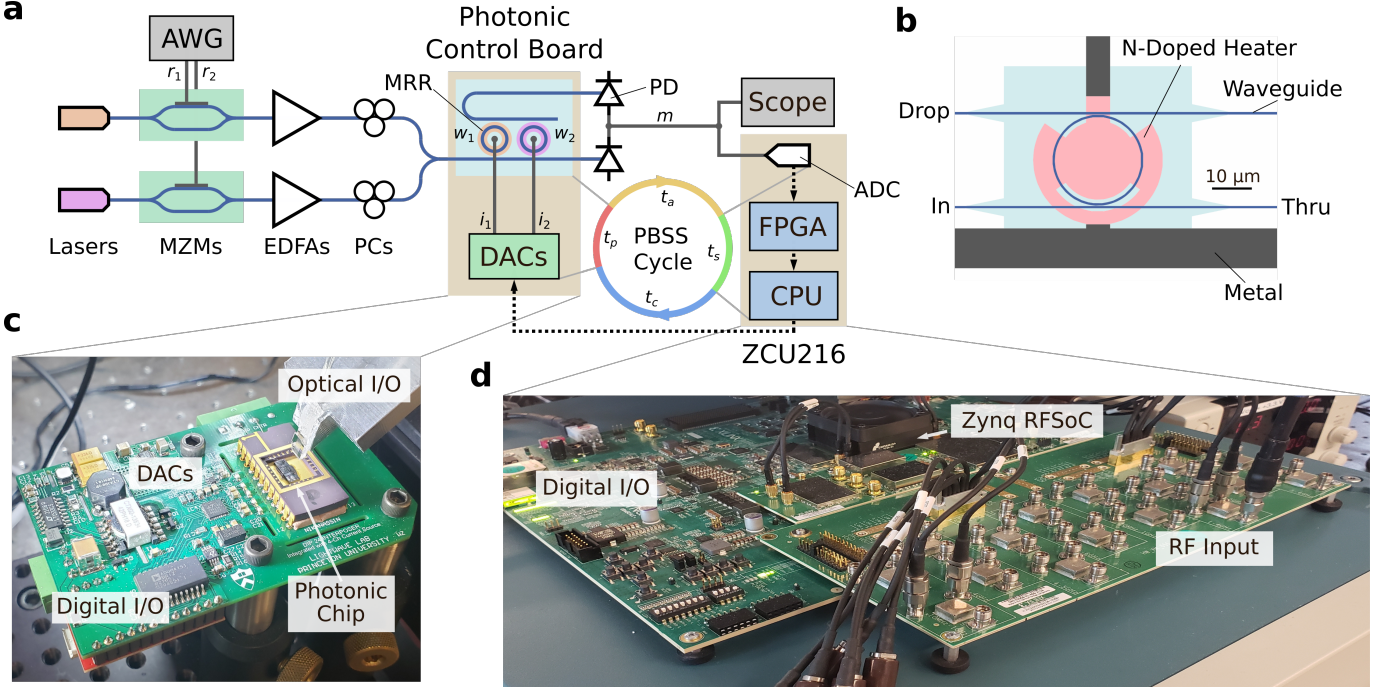} 
	\caption{Experimental blind interference cancellation setup. (a) Setup diagram. (b) Scale schematic of the experimental MRRs (variation in N-doping concentration not shown). (c) Packaged photonic chip and controller. (d) ZCU216 FPGA development board. The experimental setup generates linearly mixed signals in software and uses an arbitrary waveform generator (AWG) and Mach-Zehnder Modulators (MZMs) to modulate them onto distinct wavelengths of light, replacing the beamforming circuitry and input MRR bank of the proposed receiver. An FPGA performs statistic generation and an integrated central processing unit (CPU) acts as the PBSS engine. EDFA: erbium-doped fiber amplifier, PC: polarization controller, I/O: input/output, RFSoC: RF system-on-chip.} 
	\label{fig:experiment}
\end{figure*}

\section{Results}

\subsection{A Photonic-RF mmWave MIMO Receiver }
\label{sec:hardware}

\noindent In a hybrid digital-analog beamforming receiver, each physical antenna is connected by an array of splitters, combiners, and phase shifters to a smaller number of logical antenna ports. Each logical port has an angular sensitivity that may be independently controlled via the phase-shifter array. As a result, each port provides a linear mixture of all incident source signals, with weights dependent on the beamformer tuning, the signals' angles of incidence, and the environment. These received signals are connected to a set of RF chains that perform filtering, downconversion, and digitization (\cref{fig:diagram}a) \cite{abbas_millimeter_2017, alkhateeb_2014, sohrabi_2016, kutty_beamforming_2016, zhan_2021}.

The mixing process, neglecting non-interference noise sources, may be described as:
\begin{equation}
	\vec{r}(t) = \bm{M}\vec{s}(t)
\end{equation}
where $\vec{s}(t) = \left\{ s_1(t), s_2(t), \ldots, s_n(t) \right\}$ represents the $n$ independent source signals, the independent components (ICs), which include both target and interference sources; $\vec{r}(t) = \left\{ r_1(t), r_2(t), \ldots, r_l(t) \right\}$ represents the $l$ received signals via the logical antenna ports; and $\bm{M}\in \mathbb{R}^{l\times n}$ represents the unknown mixing matrix. For each target source signal $s_i(t)$, we seek to identify the cancellation weight vector $\vec{c}_i$ that produces a demixed signal that most closely approximates the source:
\begin{equation}
	s_i(t) \approx \vec{c}_i \cdot \vec{r}(t) = \vec{c}_i^{T} \bm{M}\vec{s}(t)
\end{equation}

We propose the novel photonic receiver architecture shown in \cref{fig:diagram}b, in which this vector multiplication is performed using analog photonics. Under our approach, the received signals are modulated by broadband MRRs onto spaced wavelengths produced by a co-integrated laser frequency comb \cite{bian_3d_2020, alic_optical_2014}, generating a wavelength-division multiplexed optical signal. The multiplexed optical signal is split between multiple photonic-RF chains, each implementing vector multiplication with tunable MRRs to perform weighting and balanced photodetectors to perform summing in a broadcast-and-weight configuration \cite{tait_broadcast_2014}. Filtering, downconversion, and digitization match that of the conventional system.

\subsection{Photonic Blind Source Separation}

\noindent As a consequence of the analog cancellation in the proposed receiver, not all received signals are digitized. This constraint, in addition to the potential lack of cooperation of interference sources, prevents the use of many conventional channel estimation techniques. Instead, we apply PBSS, a proven statistical approach which does not require full knowledge of the incident signals or action by transmitters \cite{tait_2018, ma_photonic_2019, ma_2020, zhang_2021, huang_2022, zhang_2023}. PBSS consists of an iterative algorithm that tests different photonic weight vectors $\vec{w}$ and evaluates the resulting mixed signal $m(t)$:
\begin{equation}
	m(t) = \vec{w}\cdot\vec{r}(t)
	\label{eq:photonic_mixing}
\end{equation}
PBSS requires two statistical measurements of $m(t)$: variance $\sigma^2=\mathbb{E}\left[ m^2(t) \right] $ and kurtosis $\kappa=\mathbb{E}\left[ m^{4}(t) \right]/\sigma^{4}-3$ (where it is assumed $\mathbb{E}[m(t)]=0$ for a modulated RF signal).

PBSS relies on a property stemming from the principles of independent component analysis (ICA): when $\vec{w}=\vec{c}_i$ for some $i$---i.e. when the mixed output most closely matches one of the independent source signals---$\kappa$ is at a local minimum. All such local minima can be determined using principal component analysis (PCA). PCA, which involves maximization of $\sigma^2$, enables the construction of a new weight basis in which all $\vec{c}_i$ are orthogonal, ensuring they can be deterministically identified.

In the proposed receiver, a low-latency statistic generator calculates $\sigma^2$ and $\kappa$, a PBSS engine runs the $\vec{w}$ optimization algorithm, and digital-to-analog converters (DACs) control the photonic weights by tuning the MRRs (Fig. \ref{fig:diagram}b). For the experimental demonstration of real-time interference cancellation, artificial signals with a 1 GHz carrier frequency serve as the received signals (see Methods). Critically, the setup uses an FPGA chip for digital processing, with the FPGA fabric performing statistic generation and the on-chip CPU running the optimization algorithm (Fig. \ref{fig:experiment}). A custom DAC board performs micro-ring tuning on the co-packaged photonic chip.

\subsection{Real-Time PBSS}
\label{sec:algorithm}

\begin{figure*} 
	\centering
	\includegraphics[width=\linewidth]{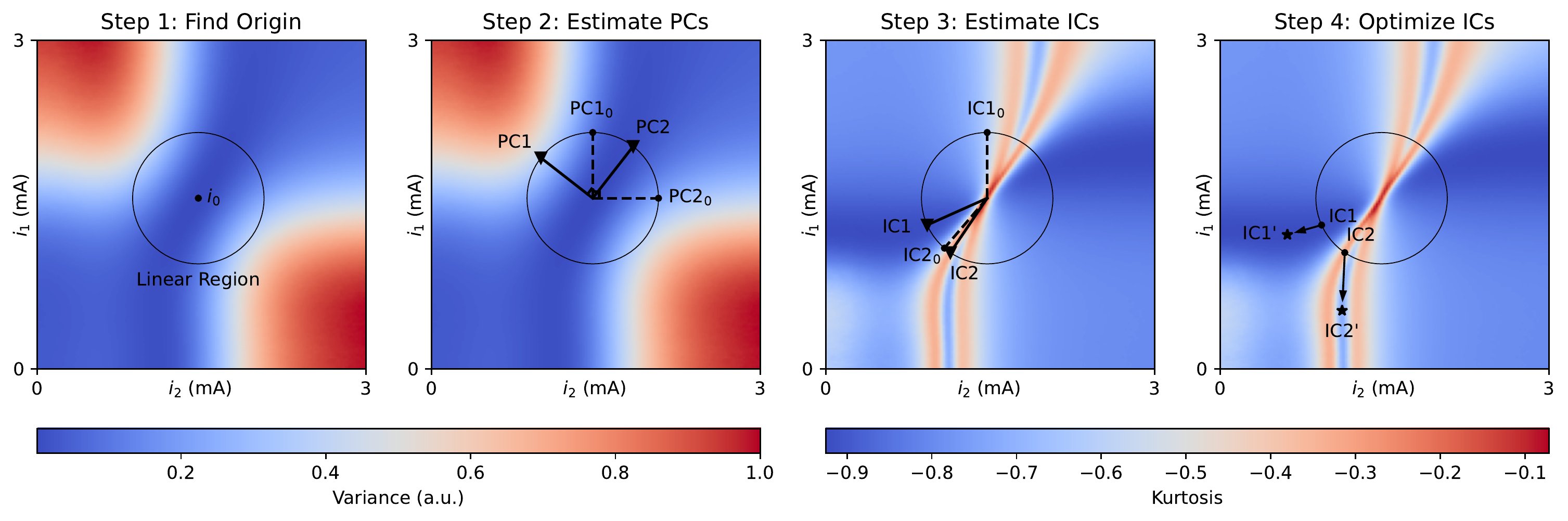}
	\caption{A typical experimental run of the PBSS algorithm with the $\bm{M}_2$ mixing matrix (Eq. \ref{eq:mixing_matrix}). Step 1 identifies the point of lowest variance, $\vec{i}_0$. Step 2 consists of PCA and step 3 of ICA, each optimizing directional vectors about the circle denoting the linear region. A 0 subscript and dashed lines denote initial positions, while solid lines denote final positions. Step 4 optimizes the ICs, freeing them from the linear region to reach higher-amplitude final positions denoted $'$. PCs: principal components.
	}
	\label{fig:sweep}
\end{figure*}

\noindent $\bm{M}$ changes in response to the movement of people and objects in the environment, but it is assumed to be static during PBSS. PBSS must therefore run to completion before $\bm{M}$ meaningfully changes, operating in real-time with low latency. Following PBSS, continuous kurtosis minimization can ensure the target weight vectors remain accurate, so the latency of the initial PBSS operation represents the limiting factor. PBSS consists of a series of weight-set/signal-measure cycles (\cref{fig:experiment}a), so PBSS latency is the product of cycle count $N$ and average cycle latency $t$. 

No prior work has achieved low-latency, real-time PBSS operation, with previous PBSS demonstrations facing latencies in the minutes, incompatible with practical application \cite{zhang_2023}. Calibration represents a particular challenge. All previously reported applications of MRR photonic systems rely on pre-calibration of the MRR weight bank to determine the nonlinear relationship between the applied weighting currents $\vec{i}$ and the weights $\vec{w}$ from Eq. \ref{eq:photonic_mixing} \cite{ma_photonic_2019, ma_2020, zhang_2023}. Calibration requires sweeps of current that add dozens of processing cycles per micro-ring, a significant latency penalty, and it may be compromised by MRR crosstalk \cite{chaoran_2020}. Furthermore, MRR characteristics can shift with operating temperature or optical input power, necessitating frequent recalibration to maintain accuracy. To avoid these challenges, we propose and demonstrate a novel zero-calibration, error-robust approach to MRR control:

For properly chosen optical wavelengths, the photonic transfer function $\vec{f}$, where $\vec{w}=\vec{f}(\vec{i})$, may be approximated as linear about the zero-weight point $\vec{i}_0$ (see Supplementary Notes): 
\begin{equation}
	\vec{f}(\vec{i}) \approx \left(\left. \mathrm{D}\vec{f}\right|_{\vec{i}_0}\right)( \vec{i}-\vec{i}_0 ) \quad \textrm{given} \quad |\vec{i}-\vec{i}_0| < i_{max}
		\label{eq:condition}
\end{equation}
where $\vec{f}(\vec{i}_0)=\vec{0}$, $\left. \mathrm{D}\vec{f}\right|_{\vec{i}_0} \in \mathbb{R}^{l\times l}$ represents the matrix of partial derivatives of $\vec{f}(\vec{i})$ evaluated at $\vec{i}_0$, and $i_{max}$ represents the maximum deviation from $\vec{i}_0$ where $\vec{f}$ may still be approximated as linear. It follows that the mixed signal $m(t)$ may be approximated as such:
\begin{equation}
	m(t) = \vec{w}^{T}\bm{M}\vec{s}(t) \approx \underbrace{(\vec{i}-\vec{i}_0)}_{\vec{w}'} {}^{T}\underbrace{\left( \left.\mathrm{D}\vec{f}\right|_{\vec{i}_0} \right)^{T} \bm{M}}_{\bm{M}'}\vec{s}(t)
\label{eq:effective}
\end{equation}
Under this approximation, there is an effective weight vector $\vec{w}'$, determined from the applied currents without calibration, and an unknown effective mixing matrix $\bm{M}'$, the product of $\left.\mathrm{D}\vec{f}\right|_{\vec{i}_0}$ and the true mixing matrix $\bm{M}$. As $\left.\mathrm{D}\vec{f}\right|_{\vec{i}_0}$ does not need to be independently determined, only $\vec{i}_0$ must be identified.

\cref{fig:sweep} shows a typical experimental PBSS run under this approach with two source and two received signals. PBSS consists of four steps. First, variance is minimized to find $\vec{i}_0$, establishing the linear region as a circle centered at $\vec{i}_0$ following Eq. \ref{eq:condition}. PCA and ICA (steps 2 and 3) operate along the edge of this circle, balancing weighting linearity with mixed signal SNR, which increases with amplitude. PCA consists of finding an orthogonal basis in which each successive basis vector points in the direction of highest variance. ICA performs an analogous process, though it operates in an adjusted basis derived from PCA and seeks to minimize kurtosis (see Supplementary Methods).

Estimates of the ICs after step 3 are subject to the cascaded errors of the previous steps stemming from noise, sampling randomness, drift in $\bm{M}$, and the linearity approximation. Furthermore, we seek to increase the amplitude of the post-cancellation signal to maximize SNR, but that requires pushing $\vec{i}$ out of the linear region. To address both problems, we add a final step to the algorithm consisting of a kurtosis minimization over the entire weight field, with each IC estimate serving as the initial position. As noise raises the kurtosis, this step optimizes both the ICs' weights and SNRs without requiring the linearity assumption of Eq. \ref{eq:condition}. This constitutes step 4. So long as steps 1-3 place the initial IC estimates within the correct convex regions, the optimizer will find the correct kurtosis minima. Experimentally, both the stronger and weaker ICs can be consistently identified and accurately demodulated under a wide range of $\bm{M}$s.

\begin{figure*}
	\centering
	\includegraphics[width=\linewidth]{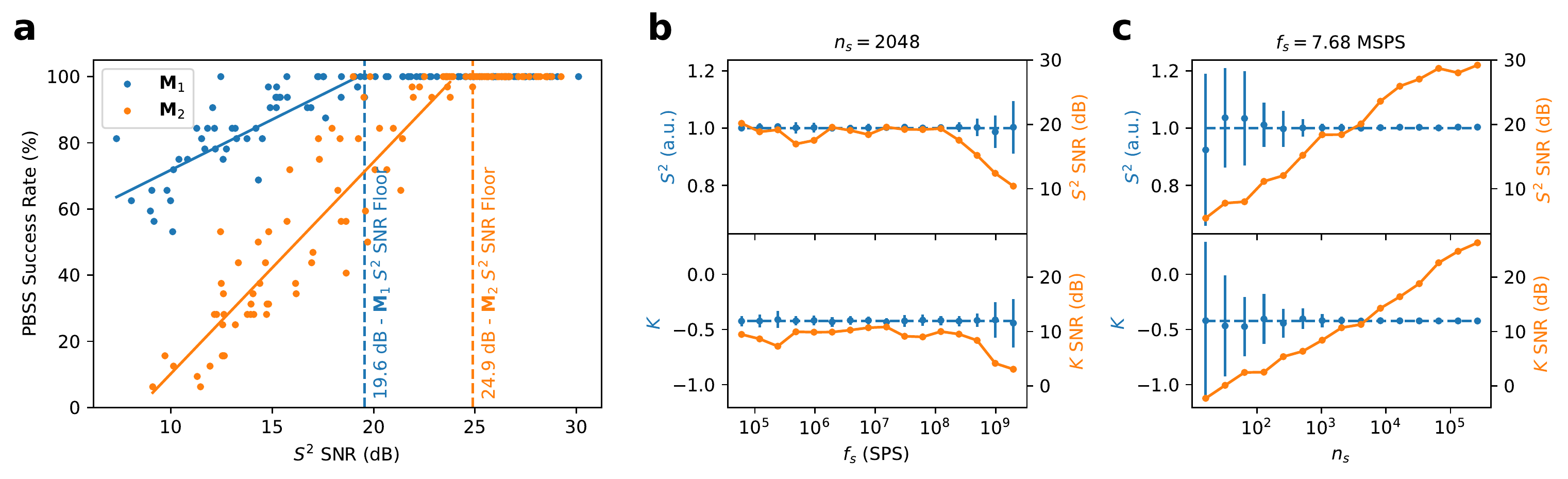}
	\caption{PBSS performance and statistic estimator consistency. (a) PBSS success rate as a function of $S^2$ SNR for $\bm{M}_1$ and $\bm{M}_2$. The solid lines show linear fits to the data with imperfect success rates, and dashed lines show the lowest SNR such that all PBSS attempts with higher SNRs were successful---the SNR floor. (b) Consistency of $S^2$ and $K$ as a function of $f_s$. (c) Consistency of $S^2$ and $K$ as a function of $n_s$. Dotted lines show the true underlying statistic values, and error bars show the statistic standard deviations. Estimator SNRs are plotted in orange. SPS: samples per second.}
	\label{fig:stats}
\end{figure*}

To achieve real-time operation, we must not just minimize $N$ but also minimize cycle latency $t$. Cycle latency consists of four non-negligible components: signal acquisition latency $t_a$, statistic calculation latency $t_s$, DAC communication latency $t_c$, and photonic weight stabilization latency $t_p$ (\cref{fig:experiment}a). Previous PBSS demonstrations were limited by a $t_s$ in the 10s of seconds due to the oscilloscope sampling \cite{zhang_2023}. The novel implementation of FPGA sampling and statistic generation in the experimental system makes $t_s$ negligible. Instead, the setup is limited by a 2-6 ms $t_c$, but it nevertheless achieves sub-second total latency. Digital interface optimization, combined with high-frequency PN-junction micro-rings, render both $t_c$ and $t_p$ negligible in the proposed system in comparison to the signal acquisition latency $t_a$ \cite{lima_2022}. $t_a$ represents the duration required to collect a sufficient number of samples of $m(t)$ to generate accurate signal statistics, and it sets the latency floor in blind cancellation systems using ICA.

\subsection{Signal Acquisition and Statistic Generation}

\noindent 
We collect $n_s$ samples of $m(t)$ at sampling rate $f_s$ to generate estimates of $\sigma^2$ and $\kappa$, denoted $S^2$ and $K$, respectively. Variation in these estimates between consecutive measurements, a result of sampling randomness, may be quantified as an estimator SNR (see Methods). As a series of statistic estimates drives the PBSS algorithm, low estimator SNRs have the potential to degrade PBSS performance. 

Fig. \ref{fig:stats}a shows the PBSS success rate as a function of $S^2$ SNR for two representative mixing matrices:
\begin{equation}
	\bm{M}_1 = \m{0.6& 0.4 \\ 0.4& 0.6} \quad \bm{M}_2 = \m{1 & 0.5 \\ 1 & 0.2}
	\label{eq:mixing_matrix}
\end{equation}
$\bm{M}_1$ represents a symmetrical case in which there is a similarly powerful interfering signal, corresponding to Fig. \ref{fig:task}a. $\bm{M}_2$ represents a case in which powerful jamming interference masks a weaker target signal, corresponding to Fig. \ref{fig:task}b.

PBSS failure typically occurs when an error in the estimate of an IC leaves it outside the often small convex region with the desired kurtosis minimum, resulting in a failure to optimize to that minimum. Variance measurements, which occur earlier in the PBSS process, have a greater ability to cause cascading errors that result in PBSS failure. Hence, $S^2$ SNR is strongly predictive of PBSS success rate, as shown in Fig. \ref{fig:stats}a. We find that PBSS is perfectly successful above a certain $S^2$ SNR floor, which depends on the mixing scenario.

The selection of $n_s$ and $f_s$ dictates both the estimator SNR for a given $\bm{M}$ and the interference canceller design. Digitization expense, in hardware and power requirements, scales with $f_s$. With mmWave signals transferring symbols at hundreds of MBaud or higher, sampling at a super-Nyquist rate becomes costly in handheld devices \cite{5G_bands}. Digitization hardware and power draw can be greatly reduced by decreasing bit precision, with precision as low as 1 bit being studied \cite{abbas_millimeter_2017, alkhateeb_2014, skrimponis_2020, heath_2016}. Such an approach is expected to become increasingly essential as carrier frequencies continue to scale. Low-precision analog-to-digital converters (ADCs) can allow effective signal demodulation, but they degrade the system's ability to determine accurate signal statistics to drive the PBSS algorithm. 

To address this limitation, we propose the dual-ADC approach for mmWave PBSS shown in Fig \ref{fig:diagram}b. The demodulation ADCs, operating at a super-Nyquist rate with low bit precision, take advantage of developments in efficient mmWave front end design. An additional sub-Nyquist ADC per photonic-RF chain with sufficient precision to allow accurate statistic generation provides data to drive the PBSS algorithm. Periodic sub-Nyquist sampling has been successfully demonstrated for PBSS, but the impact on estimator SNR and thereby PBSS success rate has not be investigated \cite{shi_2022, huang_2022}. There is significant potential for alignment between frequency components of the measured signal and the sampling rate, generating data artifacts that reduce performance. This motivates further analysis.

\cref{fig:stats}b shows estimator performance as a function of $f_s$, with $n_s$ fixed at 2,048. Estimator SNR is flat (with the exception of degradation at higher sampling rates due to an experimental artifact discussed in Supplementary Notes). The data indicates that deep sub-Nyquist sampling represents an effective approach to PBSS.

A low, fixed $f_s$ also impacts sample acquisition latency $t_a$, the primary latency component of the proposed system:
\begin{equation}
	t_a = \frac{n_s}{f_s}
\end{equation}
We seek to minimize $t_a$, and with $f_s$ limited by hardware requirements we must turn to minimizing $n_s$. Fig \ref{fig:stats}c shows the impact of $n_s$ on estimator SNR, with $f_s=7.68$ MSPS. The estimator SNRs fall sharply with $n_s$, consistent with theory. By reducing $n_s$, we compromise the PBSS success rate.

$f_s$ must be chosen in a tradeoff between hardware costs and latency, while $n_s$ trades off between latency and success rate. Specific values depend on the use case of the proposed system. One candidate set of sampling parameters, $n_s=2^{14}$ and $f_s=122.9$ MSPS, achieves 25.6 dB $S^2$ SNR, above the SNR floor for both mixing matrices tested. With $t_a=133\s{\micro\second}$, total PBSS convergence time for the proposed system operating with these parameters falls below 30 ms for two sources, suitable for all but the fastest-changing RF environments.

\section{Discussion}

\noindent We propose a photonic mmWave MIMO receiver architecture capable of performing interference cancellation without requiring downconversion and digitization of the interfering signals, a significant cost in space and power. This system has particular potential in handheld devices, where space and power are constrained. By increasing mmWave mobile devices' tolerance to interference, greater levels of multi-user spatial multiplexing become feasible, facilitating network capacity improvements \cite{alkhateeb_2014, dai_2015}.

Our experimental results represent the first demonstration of real-time applied photonic weight determination, enabled by a novel zero-calibration MRR control approach and low-latency FPGA-photonic integration. We achieve sub-second total PBSS latency, consistent with real-time operation even as $\bm{M}$ shifts due to the movement of people and objects. We further quantify the performance implications of sub-Nyquist sampling, an approach critical to practical PBSS at mmWave frequencies. We identify 30 ms as an achievable PBSS latency in the proposed system, compatible even with fast-changing RF environments, such as those in moving vehicles.

The 1 GHz experimental mixed signals, with 200 MBaud symbol rates, more accurately reflect a mmWave signal post-downconversion, rather than pre-downconversion. While downconversion prior to photonic modulation is feasible, the greatest hardware reduction may be achieved by performing cancellation directly on the received signals. MRRs have demonstrated performance up to 19.2 GHz \cite{zhang_2023}, and recent advancements in novel integrated silicon photonic components show promise toward fully extending integrated photonics to the mmWave domain \cite{weigel_bonded_2018, wang_integrated_2018, rogalski_dimensional_2019}.

We anticipate order-of-magnitude improvements in PBSS latency with the implementation of available low-latency digital communication protocols (e.g. serial peripheral interface) and high-frequency PN junction micro-rings \cite{lima_2022}. Furthermore, the number of cycles per PBSS optimization step was fixed, and dynamic adjustment of the cycles per optimization has the potential to both reduce latency by terminating optimization early and increase performance by extending or restarting optimization when convergence has not been achieved. Better characterization of performance with multiple emitters and a variety of mixing matrices is essential to validate the expected broad applicability of the proposed approach. Synergies and conflicts between the photonic linear mixer and the beamforming apparatus should also be explored.

\section{Methods}

\subsection{Experimental Setup}

\noindent The experimental source signals are two binary phase-shift keyed (BPSK) signals consisting of distinct 1,137-bit repeating random sequences. Each has a 200 MBaud data rate and a carrier frequency offset from 1 GHz by 176 kHz in opposite directions to prevent artifacts generated from a perfect alignment (the frequency offset is not used for signal discrimination). The signals are mixed in software and generated by a Keysight N8196A 92 GSPS AWG, which modulates the signals using MZMs onto distinct C-band laser frequencies generated by two Pure Photonics PPCL500 lasers. The light is polarization-controlled, amplified, and coupled onto a photonic chip, which performs linear mixing. The output intensities are received by a Discovery Semiconductor DSC-R405ER balanced photodetector, and the resulting signal is split between a Tektronix DPO73304SX 100 GSPS oscilloscope and an analog input to the Xilinx ZCU216 FPGA development board. The on-board CPU directs the system and communicates with a separate custom weight control board, which contains DACs to tune the micro-ring weights. The setup largely matches that of Zhang et. al. \cite{zhang_2023}. 

\subsection{Optimization Approach}

\begin{table}
	\centering
	\caption{Steps of the Proposed PBSS Algorithm.}
	\label{tab:steps}
	\begin{tabular}{c c c c c}
		\hline
		Step & Direction & Parameter & Domain & Repeats \\
		\hline
		1 & Minimize & Variance & Field & 1 \\
		2 & Maximize & Variance & Hypersphere & $a-1$ \\
		3 & Minimize & Kurtosis & Hypersphere & $a-1$ \\
		4 & Minimize & Kurtosis & Field & $a$ \\
		\hline
	\end{tabular}
\end{table}

\noindent The PBSS algorithm consists of the optimizations listed in Table \ref{tab:steps}, performed using the Nelder-Mead derivative-free algorithm with a fixed forty iterations per optimization \cite{nelder_mead}. $a$ represents the number of distinguishable independent signals. The optimization domain for PCA and ICA consists of a hypersphere centered at $\vec{i}_0$ with a 0.6 mA radius, corresponding to the boundary of the linear region. Each component identified constrains the optimization domain for later components under the requirement that later components are orthogonal to earlier components in the original (PCA) or whitened (ICA) basis. Hence the hypersphere dimensionality reduces by one dimension for each optimization. 

\subsection{Statistic Generation}

\noindent $n_s$ samples of $m(t)$, denoted $m_1, m_2, \ldots, m_{n_s}$, are used to generate estimators $S^{2}$ and $K$ of the variance and kurtosis, respectively:
\begin{equation}
	S^{2} = \frac{1}{n_s} \sum_{i=1}^{n_s} m_i^{2}  \quad K=\frac{1}{S^{4}}\frac{1}{n_s}\sum_{i=1}^{n_s} m_i^{4} - 3
	\label{eq:statistics}
\end{equation}
The quality of these estimators for the purpose of PBSS is quantified by their SNR. The SNR of an estimator $A$ is defined as the ratio of the estimator mean $\mu(A)=\mathbb{E}[A]$ to the estimator standard deviation $\sigma(A)=\sqrt{\mathbb{E}\left[ \left( A-\mu(A) \right)^2 \right] } $.

\subsection{Data Collection}

\noindent Estimator SNR and PBSS success rate data, shown in \cref{fig:stats}, were collected under a fixed set of sampling rates $f_s$ and sample counts $n_s$. $f_s$ varied from 960 kHz to 1.966 GHz by power-of-2 scaling factors, and $n_s$ varied from $2^{8}$ to $2^{16}$ by powers of 2. PBSS success rate is defined as the percentage of PBSS attempts, out of 32, which enable the successful demodulation of both source signals with no bit errors over the full bit sequence. Tests on all combinations of allowed $f_s$ and $n_s$ values are shown in \cref{fig:stats}a. $S^2$ and $K$ SNR data shown in \cref{fig:stats}b-c was collected with the $\bm{M}_1$ mixing matrix at the high-variance $\vec{i}=(0\s{\milli\ampere},\ 3\s{\milli\ampere})$ point, with estimator mean and standard deviations calculated from 32 statistic measurements.

\section{Data Availability}

\noindent The data generated for this work have been deposited in the Figshare database and are are available at https://doi.org/10.6084/m9.figshare.22773689.

\section{Code Availability}

\noindent All code used in this study is available from the corresponding author upon request.

\section{Acknowledgements}


\noindent This research is supported by the National Science Foundation (NSF) (ECCS-2128616 to P. Prucnal and ECCS-1642962 to P. Prucnal), the Office of Naval Research (ONR) (N00014-18-1-2297 to P. Prucnal and N00014-20-1-2664 to P. Prucnal), and the Defense Advanced Research Projects Agency (HR00111990049 to P. Prucnal). The devices were fabricated at the Advanced Micro Foundry (AMF) in Singapore through the support of CMC Micro-systems. J. C. Lederman acknowledges support from the Department of Defense (DoD) through the National Defense Science \& Engineering (NDSEG) Fellowship Program. B. J. Shastri acknowledges support from the Natural Sciences and Engineering Research Council of Canada (NSERC). S. Bilodeau acknowledges funding from the Fonds de recherche du Québec - Nature et technologies.

\section{Author Contributions}

\noindent J.C.L., W.Z., and T.F.L. conceived the idea for the experiment. J.C.L. conceived the application of PBSS to MIMO interference cancellation and developed the proposed design and the zero-calibration MRR control scheme. W.Z. developed the experimental photonic setup, including the DAC control board and the associated control software. J.C.L. integrated the ZCU216 and wrote the associated software, including the FPGA fabric design and the implementation of the PBSS algorithm, with support from T.F.L. and W.Z. T.F.L., W.Z., S.B., and E.C.B provided theoretical and experimental support. J.C.L wrote the manuscript with support from T.F.L., B.J.S., and W.Z. P.P. supervised the research and contributed to the vision and execution of the experiment. 

\section{Competing Interests}

\noindent The authors declare no competing interests.

\section{Materials \& Correspondence}

\noindent Correspondence and material requests should be addressed to J. C. Lederman.

\bibliographystyle{IEEEtran}
\bibliography{references}

\end{document}


\maketitle

\section*{Supplementary Notes}
\subsection*{The Photonic Transfer Function}
\label{sec:transfer_function}

\noindent The transfer function $T$ of a micro-ring resonator with Q-factor $Q$ and center frequency $\omega_0$ may accurately be approximated as a Lorentzian:
\begin{equation}
	T\left( \delta \right) = \frac{1}{1+\delta^2} \quad \delta=\frac{Q}{\omega_0}\left( \omega-\omega_0 \right) 
	\label{eq:mrr_model}
\end{equation}
Weight setting is performed by applying a current $i$ to a resistive heater near the micro-ring, producing a temperature shift $\Delta T$:
\begin{equation}
	 \Delta T \propto i^2 \to \Delta T = \alpha i^2
	 \label{eq:heater_model}
\end{equation}
Notably, the specific nonlinear relationship of temperature to current aids the nonlinearity, and the properties of PN-junction micro-rings would vary.

The temperature shift induces a slight index of refraction shift due to the thermo-optic effect in silicon, which in turn slightly shifts $\omega_0$. Each of these shifts are sufficiently small that they associated functions may be approximated as linear. With $\omega_{00}$ represented the center frequency at the ambient temperature:
\begin{equation}
	\omega_0 \approx \omega_{00} + \beta \Delta T \quad \beta=\frac{d \omega_0}{d T} 
	\label{eq:tuning_model}
\end{equation}

Recognizing $\left| \omega-\omega_0 \right| \ll \omega_0$ and $\left| \omega - \omega_{00} \right| \ll \omega_{00}$, we combine equations \ref{eq:mrr_model}, \ref{eq:heater_model}, and \ref{eq:tuning_model}:
\begin{equation}
	\delta \approx \frac{Q}{\omega_{00}}\left( \omega - \omega_0 \right) = \frac{Q}{\omega_{00}}\left( \omega -  \omega_{00} + \alpha\beta i^2 \right)  = a + bi^2
\end{equation}
\begin{equation}
	a \equiv \frac{Q}{\omega_{00}}\left( \omega-\omega_{00} \right)  \quad b\equiv\frac{Q}{\omega_{00}}\alpha\beta
\end{equation}
\begin{equation}
	T(i) = \frac{1}{1 + \left( a+bi^2 \right) ^2}
\end{equation}

Under an idealized model of the balanced broadcast-and-weight filter, all transmitted light generates negative photocurrent with responsivity $R$, and all other light generates positive photocurrent with the same responsivity. Hence, total photocurrent $I$ follows:
\begin{equation}
	I(i) = R\left( -T(i) + (1-T(i)) \right) = R\left( 1-2T(i) \right) 
	\label{eq:transfer_function}
\end{equation}
Equation \ref{eq:transfer_function} represents the photonic transfer function.

We seek to examine the linearity of the transfer function about $I(i)=0$. The second derivative $I''(i)$ represents a simple proxy for linearity. The second-derivative formula is fairly complex, but it simplifies greatly when evaluated at $i_0$:
\begin{equation}
	I(i_0)\equiv 0  \quad i_0 = \sqrt{\frac{1-a}{b}}  \quad I''(i_0)=2R(2a-1)b
\end{equation}

While $R$ and $b$ are fixed by the photonic system design, $a$ may be tuned by selecting the appropriate $\omega$. Selecting $a$ to maximize the linearity at the zero-weight point:
\begin{gather}
	I''(i_0)=0\to a = \frac{1}{2}\to \\ \omega = \omega_{00}\left( 1 + \frac{1}{2Q} \right) = \omega_{00} + \frac{1}{4} \omega_{\mathrm{FWHM}}
\end{gather}
Where $\omega_{\mathrm{FWHM}}$ represents the full-width at half-maximum of the transfer function $T$. The optimal operating frequency for micro-ring linearity about the zero-weight point is slightly offset from the resonance peak, a more conventional operating point. As micro-rings typically have narrow resonances and high quality factors, most tunable lasers have a range easily sufficient to set the operating wavelength at the target point.

\begin{figure}
	\centering
	\subfloat[\label{fig:mrr_tuning_variance}]{%
		\includegraphics[width=0.6\linewidth]{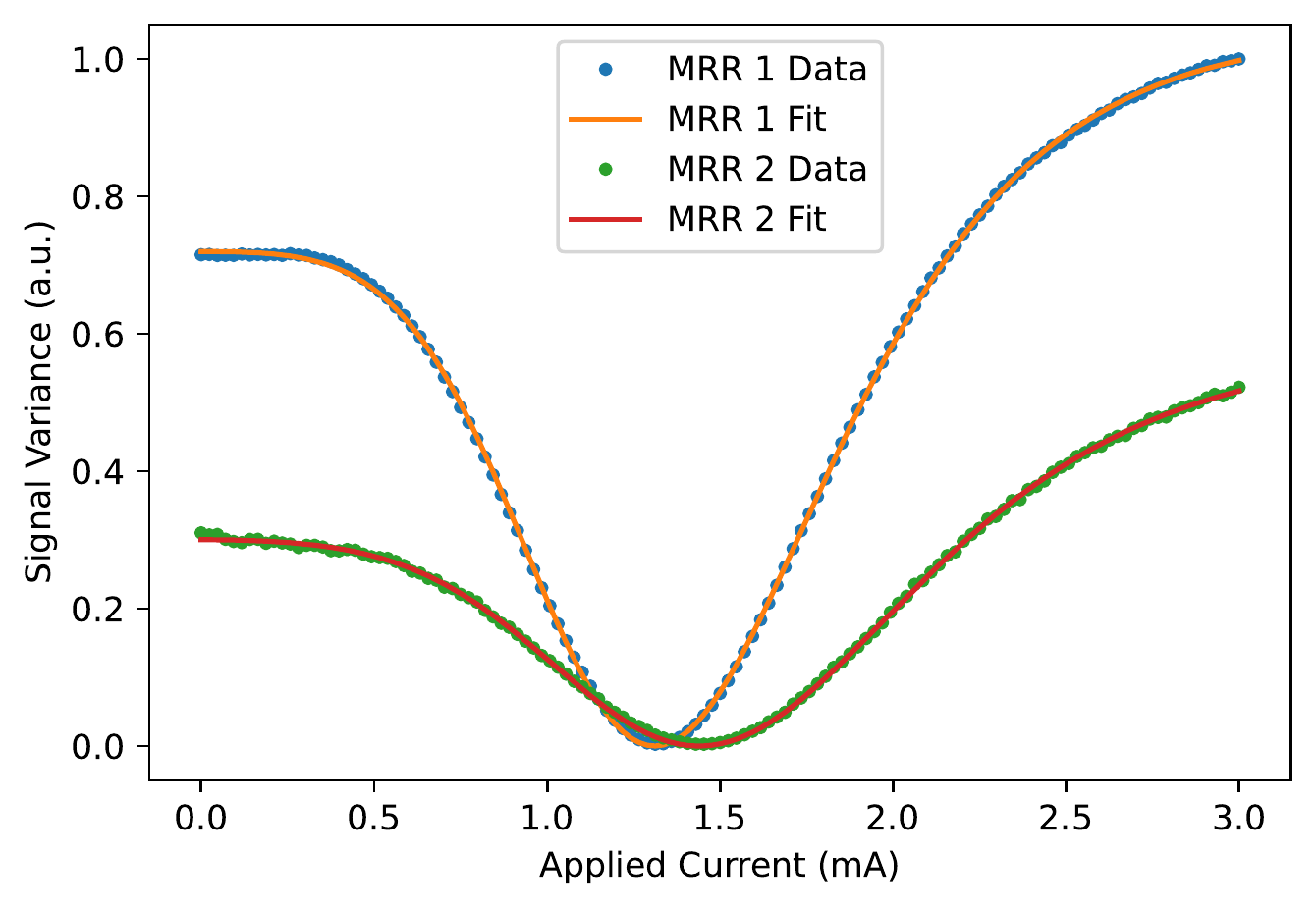}
	}
	\\
	\subfloat[\label{fig:mrr_tuning_weight}]{%
		\includegraphics[width=0.6\linewidth]{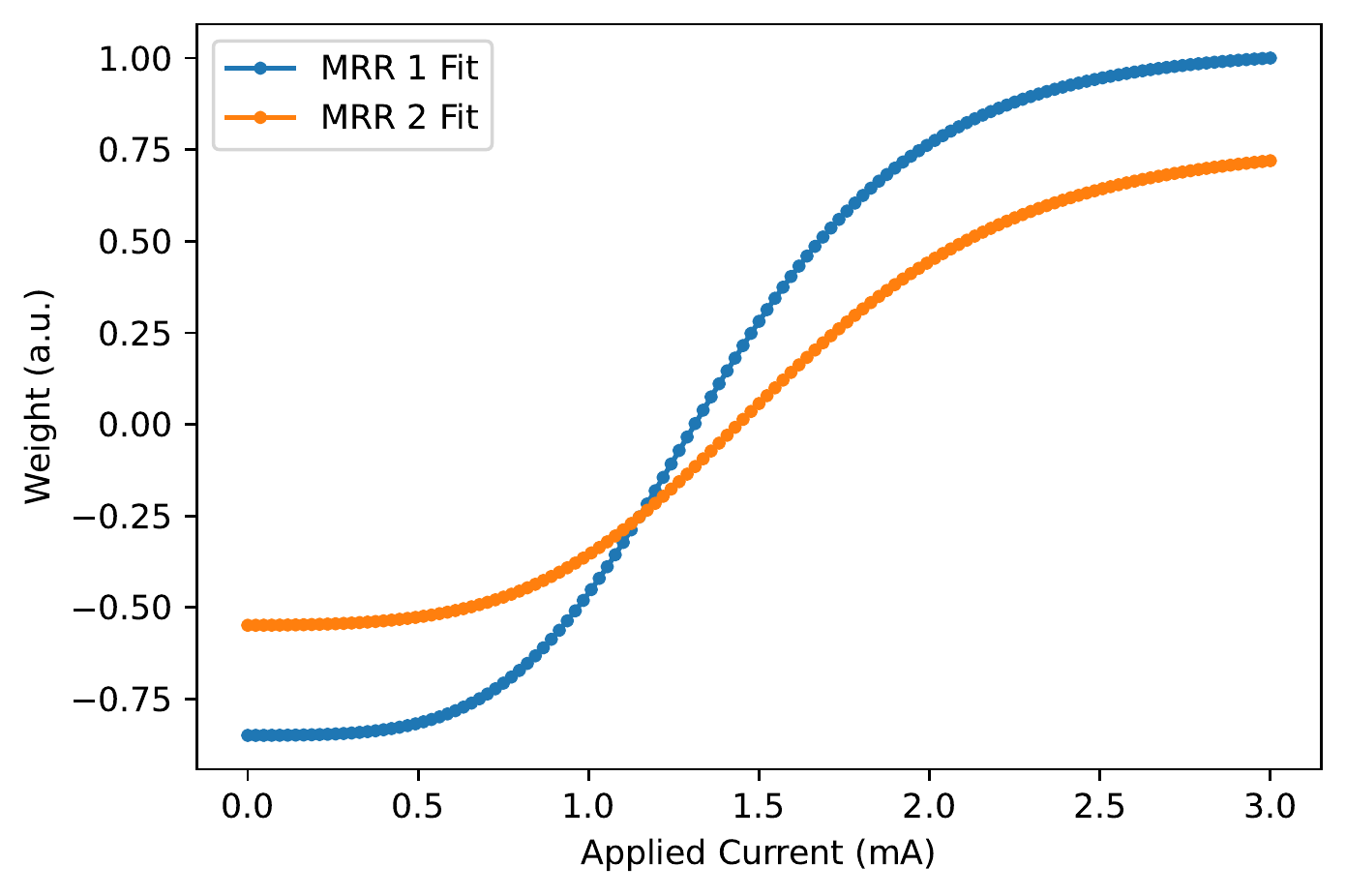}
	}
	\caption{Measured data and fitted models of the transfer functions of the two experimental MRRs. (a) Measured signal variance as tuning current is swept for each MRR, with the other MRR tuned to apply zero weight. A theoretical fit to the data is shown. (b) Fitted transfer function of each MRR, describing the relationship between applied current and effective weight. The signal variance is proportional to the square of the applied weight. Data was collected with the $\bm{M}_1$ mixing matrix using the experimental signals.}
	\label{fig:mrr_tuning}
\end{figure}

In practice, the photonic transfer function is more complex, with additional terms corresponding to micro-ring absorption, differing photodetector responsivities, and the dependence of heater resistance on temperature. Nevertheless, experimental measurements of the photonic transfer function confirm the high level of linearity at the zero-weight point, as shown in Fig. \ref{fig:mrr_tuning_weight}.

\subsection*{Derivation of Variance Minimum at Origin}
\label{sec:origin_proof}

\noindent The variance of a measured signal $m(t)$ is defined as follows:
\begin{equation}
	\sigma^2 = \int_{-\infty}^{\infty} m^2(t)dt
\end{equation}
The measured signal is a function of the source signals $\vec{s}(t)$, weights $\vec{w}$, and mixing matrix $\bm{M}$:
\begin{equation}
	m(t)=\vec{w}\cdot \vec{r}(t)  \quad \vec{r}(t)=\bm{M}\vec{s}(t)
\end{equation}
The source signals are independent, and therefore have a unit covariance matrix. With the normalization factors pulled into the matrix $\bm{M}$ for the duration of this proof:
\begin{equation}
	\int_{-\infty}^{\infty} s_i(t)s_j(t)dt = \delta_{ij} 
\end{equation}
We assume distinct mixing proportions for each signal mixture---that the mixing matrix $\bm{M}$ is linearly independent:
\begin{equation}
	\left| \bm{M} \right| \ne 0
\end{equation}
All integrals below are implied to have infinite bounds.

Calculating the gradient of $\sigma^2$:
\begin{align}
	\nabla_{\vec{w}} \sigma^2 
	&= \nabla_{\vec{w}} \int \left( \vec{w}\cdot \vec{r}(t) \right) ^2dt \\
	&= \int \nabla_{\vec{w}}\left( \vec{w}\cdot \vec{r}(t) \right) ^2dt \\
	&= 2\int \left( \vec{w}\cdot \vec{r}(t) \right) \vec{r} dt 
\end{align}
Shifting to subscript-summation notation:
\begin{align}
	\frac{\partial \sigma^2}{\partial w_{j}} &= 2 \int w_{i}r_{i}(t)r_{j}(t)dt \\
											 &= 2 \int w_iM_{ik}s_k(t)M_{jl}s_l(t)dt \\
											 &= 2 w_iM_{ik}M_{jl} \int s_k(t)s_l(t)dt \\
											 &= 2 w_iM_{ik}M_{jl} \delta_{kl} \\
	&= 2 w_iM_{il}M_{jl} \to \\
	\nabla_{\vec{w}} \sigma^2&=2\bm{M}\bm{M}^{T}w
\end{align}
Optima of $\sigma^2$ with respect to $\vec{w}$ may only exist when the gradient is zero:
\begin{equation}
	\nabla_{\vec{w}}\sigma^2 = 2\bm{M}\bm{M}^{T}\vec{w}=\vec{0}
\end{equation}
\begin{equation}
	\left| \bm{M}\bm{M}^{T} \right| =0 \quad \mathrm{or}  \quad \vec{w}=\vec{0}
\end{equation}
\begin{equation}
	\left| \bm{M} \right| \ne 0 \to \left| \bm{M} \right| ^2=\left| \bm{M}\bm{M}^{T} \right| \ne 0\to \vec{w}=\vec{0}
\end{equation}
Optima may only exist at the single point where $\vec{w}=\vec{0}$. Furthermore:
\begin{equation}
	\frac{\partial \sigma^2}{\partial w_i \partial w_j} = 2M_{ik}M_{jk} \to \bm{H}_{\sigma^2} = 2MM^{T}
\end{equation}
Where $\bm{H}_{\sigma^2}$ represents the Hessian matrix of $\sigma^2$ with respect to $\vec{w}$.
\begin{equation}
	\left| \bm{H}_{\sigma^2} \right| = 2\left| \bm{M} \right| ^2 \quad \left| \bm{M} \right| \ne 0 \to \left| \bm{H}_{\sigma^2} \right| > 0
\end{equation}
The variance is uniformly convex. Therefore there exists a single local minimum of $\sigma^2$ at $\vec{w}=\vec{0}$.

\subsection*{Statistic Mathematics}
\label{sec:statistic_mathematics}


\noindent The variance $\gamma^2$ of the variance estimator $S^2$ may be defined:
\begin{equation}
	\gamma ^2=E\left[ \left( S^2-E\left[ S^2 \right]  \right) ^2 \right] 
\end{equation}
It may be approximated with $m\gg 1$ samples of $S^2$:
\begin{equation}
	\gamma^2 \approx \frac{1}{m-1}\sum_{i=1}^{m}\left( S_i^2 - \frac{1}{m}\sum_{j=1}^{m} S_j^2  \right)^2
	\label{eq:varvar_sampling}
\end{equation}
If the signal samples were randomly chosen in time and therefore independent and identically distributed (IID), there would be a fixed relationship between the variance of the signal variance and the underlying signal variance and kurtosis:
\begin{equation}
	\gamma^2 = \frac{\sigma^{4}}{n_s}\left( \kappa-1+\frac{2}{n_s-1} \right) \approx \frac{S^{4}}{n_s}\left( K-1+\frac{2}{n_s-1} \right) 
	\label{eq:varvar_iid}
\end{equation}
Where $n_s$ represents the number of samples used to calculate the estimator. Note that for large $n$:
\begin{equation}
	\gamma \propto \frac{1}{\sqrt{n_s} }
	\label{eq:prop}
\end{equation}
Though the samples are periodic and therefore not IID, Eq. \ref{eq:prop} nevertheless describes in general terms the increase in measurement consistency observed as $n$ increases.

\begin{figure}
	\centering
	\includegraphics[width=0.6\linewidth]{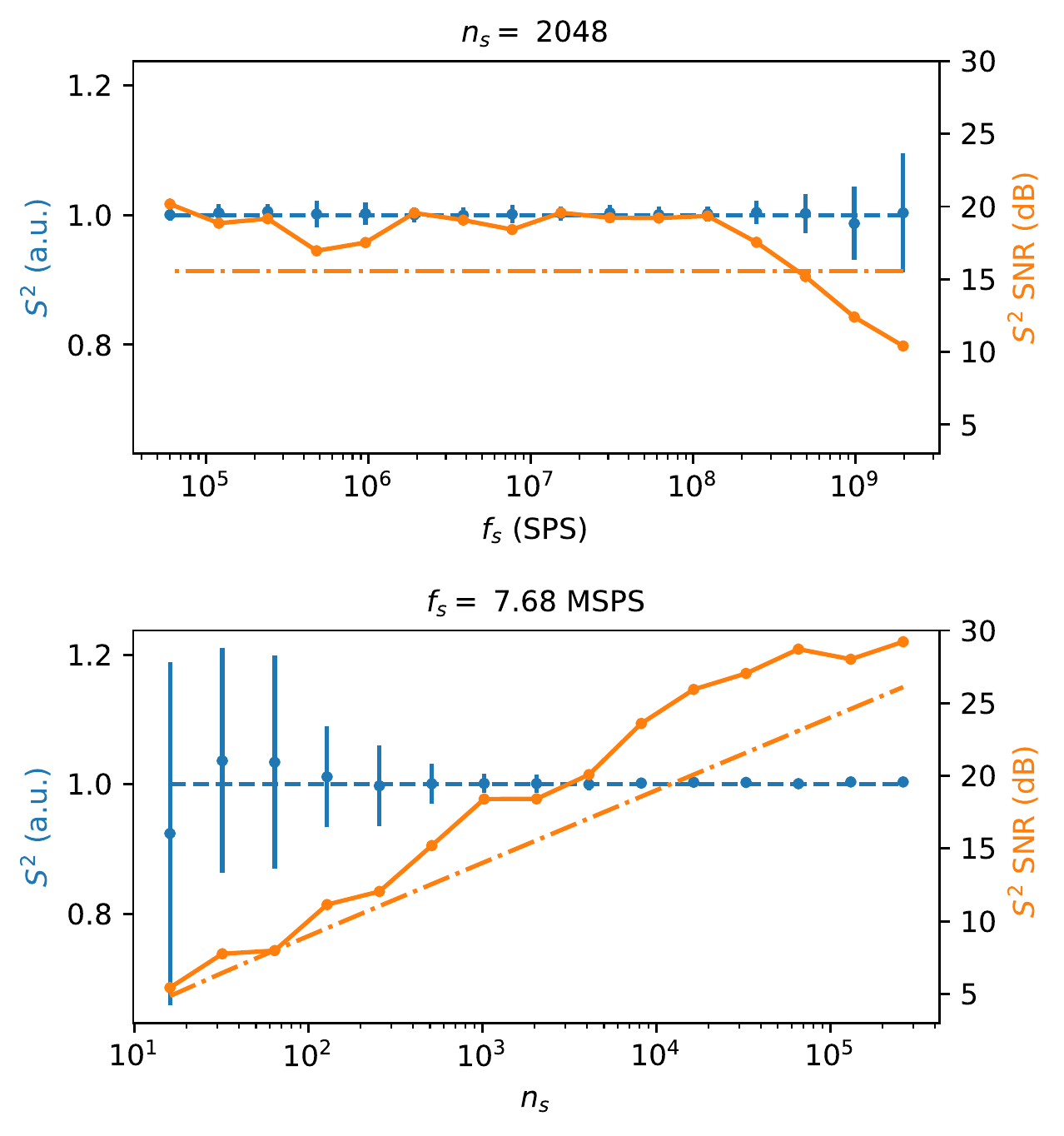}
	\caption{$S^2$ consistency measurements. This data matches that shown in Fig. 5b-c with the addition of dot-dash orange lines representing the theoretical $S^2$ SNRs expected if the collected samples were randomly chosen, following to Eq. \ref{eq:varvar_iid}.}
	\label{fig:stats_iid}
\end{figure}

Fig. \ref{fig:stats_iid} shows the variance estimator data of Fig. 5b-c with the addition of predicted estimator SNRs according to Eq. \ref{eq:varvar_iid}, assuming random sampling. Under most sampling parameters periodic sampling meaningfully outperforms random sampling, achieving a greater level of statistic measurement consistency. This is likely owed to the more even distribution over time of periodic samples than random ones. However, periodic sampling may produce very poor behavior when the sampling rate matches too closely to underlying signal frequency components. Samples would be chosen selectively from some parts of the signal distribution depending on the exact sampling start time, greatly increasing the measurement variability. Moreover, periodic sampling also performs poorly when there are signal frequency components with a period comparable to that of the total sampling duration. Sampling that occurs predominantly during a fast-changing portion of the signal (i.e. near the zero-point of a sinusoid) would measure distinct statistics as compared to sampling occurring during a slow-changing portion of the signal (i.e. near the peak of a sinusoid). Care must be taken to choose sampling durations that are best suited to the RF environment.

During experimental measurements a 1,137-bit random repeating 200 MBaud RF signal was generated for separation, with a length limited by the AWG storage capacity. This 5.68\s{\micro\second} periodic signal degrades statistic measurement consistency where the measurement time has a comparable period. When taking 2,048 samples per measurement this artifact appears at high sampling rates between approximately 500 MSPS and 2 GSPS, accounting for the sharp drop in performance of periodic sampling in comparison to random sampling at this portion of Fig. \ref{fig:stats_iid}. When sampling at 7.68 MSPS this artifact occurs at a very low samples per measurement and is therefore obscured by the naturally higher variability of such measurements. The kurtosis data shows similar behavior and may be explained in the same manner. Such artifacts result from the experimental setup and would not be observed using aperiodic bitstream data.

\clearpage
\section*{Supplementary Methods}

\subsection*{ICA and PCA for PBSS}
\label{sec:ica_pca}

\noindent Independent Component Analysis (ICA) and Principal Component Analysis (PCA) represents the primary tools used to perform PBSS.

ICA relies on the fact that, according to the Central Limit Theorem, linear mixtures of independent signals will, in general, have a distribution of values that more closely matches a Gaussian distribution than the source signals. By maximizing the non-Gaussianity of a mixture, one can identify the underlying sources of that mixture.

One straightforward measure of Gaussianity is excess kurtosis, referred to here as the kurtosis. The kurtosis of a Gaussian distribution is exactly 0, while other distributions may have higher or lower kurtoses. Sinusoids, in particular, have a kurtosis of -1.5, and it follows that RF signals, which are typically constructed of sine-like constituents, tend to have below-Gaussian kurtoses. Therefore, choosing $\vec{w}$ to minimize the kurtosis of $m(t)$ ensures $m(t)$ corresponds to one of the source signals $s_i(t)$.

There are multiple $\vec{w}$ corresponding to local kurtosis minima, each one associated with one source signal. Standard mathematical minimization techniques can find one such weight vector in the vast majority of cases, but identifying all local kurtosis minima, regardless of their position, remains a challenge. For this we turn to PCA.
In this formulation, the principal components of $m(t)$ with respect to $\vec{w}$, labeled $\vec{w}_{\mathrm{PC}i}$, correspond to the directions of maximum variance of $m(t)$, with the constraint that each subsequent component must be orthogonal to all previous components:
\begin{equation}
	\begin{aligned}
		\vec{w}_{\mathrm{PC1}} &=\argmax_{\vec{w};\ \left\Vert \vec{w} \right\Vert = 1} \sigma^2 \\
		\vec{w}_{\mathrm{PC2}} &= \argmax_{\vec{w};\ \left\Vert \vec{w} \right\Vert=1;\ \vec{w}\perp \vec{w}_{\mathrm{PC1}} } \sigma^2 \\
					  &\quad\vdots
	\end{aligned}
\end{equation}

The principal components and their associated variances may be applied to construct a \emph{whitening matrix} $\bm{W}_\mathrm{PC}$. The whitening matrix describes a transformation to a new weight coordinate basis, denoted with $'$:
\begin{equation}
	\vec{w}' = \bm{W}_\mathrm{PC}^{-1}\vec{w}
\end{equation}
$m(t)$ in the new basis has a unique property: weight vectors corresponding to the independent source signals are orthogonal. We can now solve the problem posed earlier; once we find one independent component by minimizing kurtosis we can then run a further kurtosis minimization under the constraint that the weight vector must be orthogonal to that of the original independent component. ICA may be completed in a process analogous to independent component extraction:
\begin{equation}
	\begin{aligned}
		\vec{w}'_{\mathrm{IC1}} &=\argmin_{\vec{w}';\ \left\Vert \vec{w}' \right\Vert = 1} \kappa \\
		\vec{w}'_{\mathrm{IC2}} &= \argmin_{\vec{w}';\ \left\Vert \vec{w}' \right\Vert=1;\ \vec{w}'\perp \vec{w}'_{\mathrm{IC1}} } \kappa \\
					  &\quad\vdots
	\end{aligned}
\end{equation}
The cancellation vector associated with source signal $i$, $\vec{c}_i$ follows:
\begin{equation}
	\vec{c}_i=\bm{W}_{\mathrm{PC}}\vec{w}'_{\mathrm{IC}i}
\end{equation}

\subsection*{FPGA Development}

\noindent Processing was performed by the Xilinx XCZU49DR chip with a cointegrated ADC, FPGA, and Arm cortex. The ADC sampled the measured signal continuously at a fixed rate, feeding data to the FPGA logic fabric. The statistic generator was implemented on the FPGA, accumulating data points and calculating the resulting variance and kurtosis. The Arm cortex requested statistic generation as necessary and accessed the resulting signal statistics. It further executed the PBSS algorithm, interfacing with the DAC control board over a serial protocol.

The statistic generation system on the FPGA was designed using the Xilinx High-Level Synthesis (HLS) toolchain. Sampling occurred at a fixed rate, and samples were dropped on a periodic basis as necessary to simulate lower sampling rates. As the signal mean could be assumed to be zero, the contributions of each sample $m$ to the variance and kurtosis sums---$m^2$ and $m^{4}$, respectively---could be continuously accumulated as each new sample became available. This enabled an efficient pipelined architecture that contributes negligible latency to the system, as statistic generation occurred in parallel with signal acquisition. Once the specified number of samples had been collected, the variance and kurtosis sums were divided by the number of samples to calculate the second moment (the variance) and the fourth moment. The system only allowed sample counts that are factors of two, allowing this division to be implemented with highly efficient right-shifting. The fourth moment must be divided by the square of the variance to determine the kurtosis, a calculation implemented on a floating-point basis by the Arm cortex.

\bibliographystyle{IEEEtran}
\bibliography{references}